\documentstyle[psfig,epsfig,12 pt]{article}
\def\ve{\varepsilon}

\def\sss{\scriptscriptstyle}

\newcommand{\DB}[1]{\Delta^{\!(#1)}}
\newcommand{\DF}[1]{{\cal D}^{(#1)}}

\newcommand{\Dp}[1]{D^{(#1)}}
\renewcommand{\S}[1]{S^{(#1/2)}}

\def\TT{{\rm TT}}

\def\psib{\overline\psi\hspace{-2.6mm}\phantom{\psi}}

\def\ve{\varepsilon}
\def\D{{\cal D}}

\def\r{\rho}
\def\s{\sigma}

\def\wh#1{\widehat{#1}}
\def\wt#1{\widetilde{#1}}

\def\ol#1{\overline{#1}}
\def\sss{\scriptscriptstyle}

\def\d{\partial}
\def\m{\mu}
\def\n{\nu}

\def\ul{\underline}

\def\be{\begin{equation}}
\def\ee{\end{equation}}
\def\beq{\begin{equation}}
\def\eeq{\end{equation}}
\def\bea{\begin{eqnarray}}
\def\eea{\end{eqnarray}}
\def\beqa{\begin{equation}\begin{array}{l}}
\def\eeqa{\end{array}\end{equation}}
\def\eqn#1{(\ref{#1})}
\def\eqref#1{eq.~(\ref{eq:#1})}

\def\a{\alpha}
  \def\g{\gamma}

\def\L{{\it\Lambda}}

\def\w{\omega}

\def\nn{\nonumber}

\begin{document}
\thispagestyle{empty} \setcounter{page}{0}
\renewcommand{\theequation}{\thesection.\arabic{equation}}

{\hfill{BRX TH-498}}

{\hfill{\tt hep-th/0202053}}

\vspace{2cm}

\begin{center}
{\bf GAUGE INVARIANCE WITH MASS:\\
HIGHER SPINS IN COSMOLOGICAL SPACES}

\vspace{1.4cm}

S. DESER\footnote{Invited lecture, Francqui Conference, ``Strings
and Gravity: Tying the Forces Together", Brussels, October 2001.}\\

\vspace{.2cm}

{\em Department of Physics, Brandeis University} \\
{\em Waltham, MA 02454, USA} \\
\end{center}

\vspace{-.1cm}

\centerline{{\tt deser@brandeis.edu}}

\vspace{1cm}

\centerline{ABSTRACT}

\vspace{- 4 mm}

\begin{quote}\small
I review recent work on massive higher ($s>1$) spins in constant
curvature (deSitter) spaces.  Some of the novel properties that
emerge are:  partial masslessness and new local gauge invariances,
unitarily forbidden ranges of mass, correlation between
fermions/bosons and nagative/positive cosmological constant $\L$
and finally the consistency requirement that in the limit of
infinite spin towers, $\L$ must tend to zero.
\end{quote}

\baselineskip18pt

\newpage

\section{Introduction}

This is a report on a program being carried out in collaboration
with A.\ Waldron \cite{001}, concerning the behavior of fields in
constant curvature backgrounds.  Perhaps the main novelty is that
this is the first place that major and dramatic kinematical
effects (as against the usual dynamical one) of gravity on matter
are encountered.

Although one of our initial motivations was provided by a recent
revival in the context of $\L \neq 0$ of the famous vDVZ paradox,
that the $m\rightarrow 0$ limit of massive spin 2 \cite{002} as
well as $s=3/2$ \cite{003} fields in flat space leads to a finite
discontinuity (compared to $m\equiv 0$) in the source interactions
they mediate, most of this report will not deal with that issue.
What happens, in both $s=2$ \cite{004} and $s=3/2$ [first paper of
ref. \cite{001}], is that the limits of $(m^2, \L ) \rightarrow
(0,0)$ can (depending on the path) yield almost any desired value.
[There is also an amusing Newtonian limit aspect to this question,
discussed by B.\ Tekin and myself \cite{005} that I will briefly
cover at the end.] To us the importance of the observations is
that they are just a corner of the wider truth that the physics of
massive higher spin $(s>1)$ fields (in contrast to $s\leq 1$) is
in fact governed by a new phase plane coordinatized by $(m^2 , \L
)$, instead of the one-dimensional $m^2$ line at $\L = 0$.  This
plane displays a phase structure: it is covered by transition
lines that divide it into physical and non-unitary ``phases" for
the spin in question, and these lines are themselves
representations of entirely novel, non-Minkowskian, systems with
partial masslessness, new local gauge invariances and associated
truncated helicity count.  They show how the confluence at $\L =
0$ of the concepts of masslessness, gauge invariance (with maximal
helicity only) and null propagation, is lifted (continuously this
time) when  $\L \neq 0$.  Bosons and fermions each display these
properties, albeit in  characteristically different regions of the
$(m^2 , \L )$ plane: respectively in deSitter (dS) and Anti
deSitter (AdS). This difference between the two types of particles
actually leads to a truly ``emergent" result: the cosmological
constant, being bounded by the highest particle species spins,
must vanish for towers of bosons and fermions in the limit of
infinite spins!

While we will not include all the details to be found in
\cite{001}, we will show how the effect of (A)dS backgrounds
arises in Sec.\ 2, explain the ensuing nonunitary domains in Sec.\
3, analyze spin 2 canonically in Sec.\ 4 and higher spins in Sec.\
5.  Null propagation of the partially massive models is obtained
in Sec.\ 6, the Newtonian limit explained in Sec.\ 7 and our
``solution" of the cosmological constant problem is given in Sec.\
8.

I end this introduction with a simple Figure that summarizes many
of the above results, and should clarify much of the detailed
calculations below.

\vspace{.2in}
\begin{center}
\epsfig{file=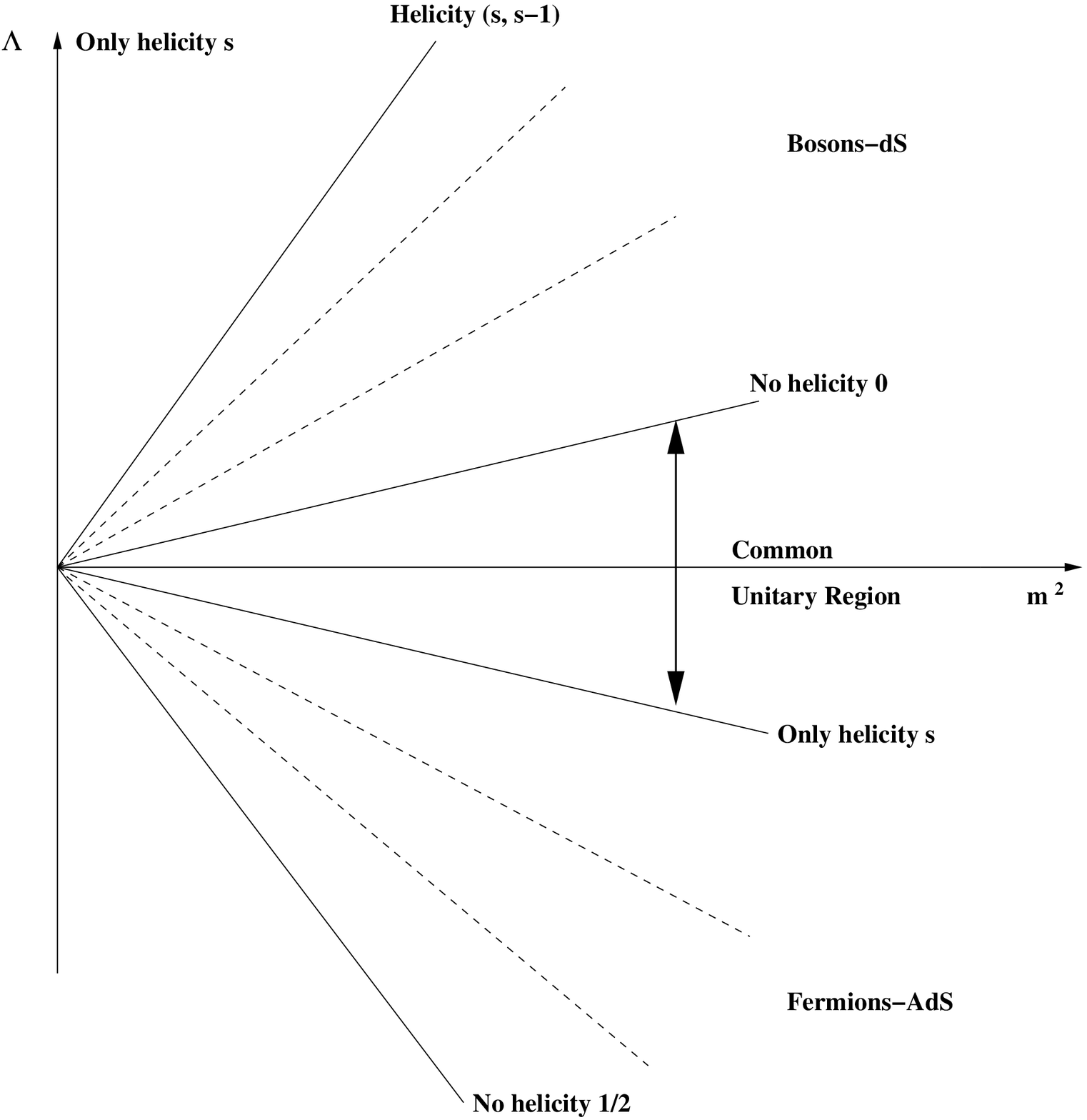, width= 10.5 cm}
\end{center}
\vspace{.2in}
\newpage 
\noindent\footnotesize{\bf Figure 1}:  The top/bottom halves of
the half-plane represent dS/AdS (and also bosons/ fermions)
respectively. The $m^2=0$ vertical is the familiar massless
helicity $\pm s$ system, while the other lines in dS represent
truncated (bosonic) multiplets of partial gauge invariance: the
lowest has no helicity zero, the next no helicities $(0,\pm 1)$,
etc.  Apart from these discrete lines, bosonic unitarity is
preserved only in the region below the lowest line, namely that
including flat space (the horizontal) and all of AdS. In the AdS
sector, it is the topmost line that represents the pure gauge
helicity $\pm s$ fermion, while the whole region below it,
including the partially massless lines, is non-unitary. Thus, for
fermions, only the region above the top line, including the flat
space horizontal and all of dS, is allowed.  Hence the overlap
between permitted regions straddles the $\L = 0$ horizontal and
shrinks down to it as the spins in the tower of spinning particles
grow; only $\L = 0$ is allowed for generic ($m^2$ not growing as
$s^2$) infinite towers.

\normalsize
%following from partialmass.tex

\renewcommand{\theequation}{2.\arabic{equation}}
\setcounter{equation}{0}

\section{Field Equations and Identities in Constant Curvature Spaces}

\label{begin}

The Riemann tensor in constant curvature spaces is \be
R_{\m\n\r\s}=-\frac{2\L}{3}\,g_{\m[\r}g_{\s]\n}\ ; \ee the
cosmological constant $\L$ is positive in dS and negative in AdS.
The actions of commutators of covariant derivatives are summarized
by the vector-spinor example\footnote{Our metric is ``mostly
plus'', Dirac matrices are ``mostly hermitean'' and the Dirac
conjugate is $\ol \psi\equiv \psi^\dagger i\g^0$. We denote
(anti)symmetrization with unit weight by round (resp. square)
brackets. Antisymmetrized products of Dirac matrices are given by
$\g^{\m_1\ldots\m_n}\equiv\g^{[\m_1}\cdots\g^{\m_n]}$.} \be
[D_\m,D_\n]\,\psi_\r=\frac{2\L}{3}\,g_{\r[\m}\psi_{\n]}
+\frac{\L}{6}\,\g_{\m\n}\psi_\r\, . \ee

The actions and field equations for massive spins (3/2,2) in
constant curvature backgrounds\footnote{For $s=2$ in generic
gravitational backgrounds, the minimally coupled Pauli--Fierz
action does not yield field equations with the correct $2s+1=5$
excitation count \cite{Boulware:1972}. Recently
\cite{Buchbinder:2000ar}, correct $s=2$ actions have been
constructed in background Einstein spaces.} are, in an obvious
notation,
 \be
{\cal L}^{(3/2)}=-\sqrt{-g}\; \psib^\m\,{\cal R}_\m\ , \;\;\;{\cal
L}^{(2)}=\frac{1}{2}\,\sqrt{-g}\;\phi^{\m\n}\; {\cal G}_{\m\n}\
,\qquad \label{ramit}
\ee
 \bea
{\cal R}_\m&\equiv&\g_{\m\n\r}\D^\n\psi^\r
=(\DF{3/2}- \gamma \cdot D)\,\psi_\m-m\,\g_{\m\n}\psi^\n\, ,\label{RS}\\
{\cal G}_{\m\n}&\equiv&
(\DB2-m^2+\L)\,(\phi_{\m\n}-g_{\m\n}\phi_\r{}^\r)+\L\phi_{\m\n}\nn\\&+&
\!D_{(\m}D_{\n)}\,\phi_\r{}^\r-2D_{(\m}D.\phi_{\n)}
+g_{\m\n}D.D.\phi\, .\label{ES}
 \eea
The operators $\DB{2}$ is the wave operators
of~\cite{Lichnerowicz:1961}
 \be
\DB2\,\phi_{\m\n} \equiv D^2\phi_{\m\n}
-\frac{8\,\L}{3}\,\Big(\phi_{\m\n}-\frac{1}{4}\,g_{\m\n}\,
\phi_\r{}^\r\Big) \ ,
 \ee
 whose introduction is justified by the
following identities, \be
\begin{array}{cc}
\DB2\, D_{(\m} \phi_{\n)} = D_{(\m}\DB1\,\phi_{\n)}\, ,&
D^\m \DB2\,\phi_{\m\n}=\DB1\,D.\phi_\n\, ,\\ \\
\DB2\,g_{\m\n}\phi=g_{\m\n}\DB0 \,\phi\, ,& g^{\m\n}\DB2
\phi_{\m\n}=\DB0\,\phi^\r{}_\r\, .
\end{array}
\label{ids}
 \ee
 The fermionic version, $\DF{3/2}$, is
given by [see first paper of ref. \cite{001}]
 \be
  \DF{3/2}\,\psi_\m \equiv\ 2\gamma \cdot
D\,\psi_\m-D_\m\g.\psi+ \g_\m(\gamma \cdot D\g.\psi-D.\psi)=\g_{\m\n\r}D^\n
\psi^\r+\gamma \cdot D\psi_\m\ ,\quad
 \ee
 and satisfies analogous identities
  \be
\begin{array}{cc}
\DF{3/2} \,D_\m\,\psi=D_\m\, \DF{1/2}\,\psi\, , &
 D^\m\,\DF{3/2}\,\psi_\m=\DF{1/2}\,D.\psi \, ,\\ \\
\DF{3/2} \,\g_\m\,\psi=\g_\m\, \DF{1/2}\,\psi\, ,&
\g^\m\,\DF{3/2}\,\psi_\m=\DF{1/2}\,\g.\psi \, .
\end{array}
\ee
 Notice that we have written the massive  Rarita--Schwinger
field equation~\eqn{RS} in terms of the $s=3/2$ operator
$\DF{3/2}$ with explicit mass term as well as in a more compact
form involving the operator \be \D_\m\equiv
D_\m+\frac{m}{2}\,\g_\m\, ,\qquad
[\D_\m,\D_\n]=[D_\m,D_\n]+(m^2/2)\,\g_{\m\n}\, , \ee encountered
in cosmological supergravity~\cite{Townsend:1977qa}.

For $s\geq1$ there are more relativistic field components than
degrees of freedom (DoF). As usual, the correct DoF count is
obtained by studying the constraints implied by divergences and
(gamma-)traces of the field equations. Less usual, for special
lines in the $(m^2,\Lambda)$ plane,  these constraints are
satisfied identically and become Bianchi identities associated
with gauge invariances. Explicitly, the divergences of the field
equations (2.4)-\eqn{ES} read
 \be
\D.{\cal R}=-\frac{1}{2}\,(3m^2+\L)\,\g.\psi\,, \;\;\; D.{\cal
G}_\n=-m^2\,(D.\phi_\n-D_\n\phi) \label{const}
 \ee
 and are constraints for generic values of the parameters $m^2$ and $\L$.
For $s=2$, the value $m^2=0$ yields Bianchi identities and their
associated (``general coordinate'') gauge invariances
 \be
 \delta \phi_{\m\n}=D_{(\m}\xi_{\n)}\, ,
 \ee
the $m^2=0$ theories are strictly massless: they propagate with
two physical helicity states. For $s=3/2$, the sole gauge
invariance
 \be \delta \psi_\m=\D_\m \varepsilon =D_\m\varepsilon+\frac{1}{2}\,
 \sqrt{-\L/3}\,\g_\m\,\varepsilon \,
, \ee
 is inherited from cosmological supergravity and occurs for
$m^2=-\L/3$ in AdS. This (rather than the $m=0$ model) is the
strictly massless helicity $\pm3/2$ theory.

For $s=2$ there is a new effect: the field equation has two open
indices and thereby also admits a double divergence Bianchi
identity. The double divergence constraint
 \be D.D.{\cal
G}+\frac{1}{2}\,m^2\,{\cal
G}_\r{}^\r=\frac{1}{2}\,m^2\,(3m^2-2\L)\,\phi_\rho{}^\rho\, ,
\label{sticky}
 \ee
  becomes a Bianchi identity not only along the
strictly massless line $m^2=0$, but also along the dS gauge line
$m^2=2\L/3$, which actually corresponds to the theory
of~\cite{Deser:1983tm}. The gauge invariance associated with this
identity,
 \be
 \delta \phi_{\m\n}=\Big(D_{(\m}D_{\n)}+\frac{\L}{3}\,
 g_{\m\n}\Big)\;\xi\, ,
 \ee
may be employed to show that this partially massless model
propagates with helicities $(\pm2,\pm1)$ on the null cone in dS.
We  discuss this theory further in Section~\ref{whale}.

To summarize, the $(m^2,\Lambda)$ half-plane offers no surprises
for spins $s\leq1$,  the usual null propagating massless theories
inhabit the line $m^2=0$ and all other points $m^2\geq0$ describe
$2s+1$ massive DoF. For $s=3/2$, the massless line is $m^2=-\L/3$
and bisects the $(m^2,\Lambda)$ half-plane. For $s=2$, the
massless theory again lies on the axis $m^2=0$ and a new gauge
invariance emerges at $m^2=2\L/3$ which also divides the
half-plane into two distinct physical regions.  This pattern will
be seen to continue for $s > 2$ as well.

\section{\hspace{-.5cm}(Anti)commutators and Nonunitary Regions}

\renewcommand{\theequation}{3.\arabic{equation}}
\setcounter{equation}{0}

\label{keepgoing}

Our analysis is rather simple and harks back to the original
inconsistency of the local field theory of charged spin~3/2
particles~\cite{Johnson:1961vt}. Generically, given
(anti)commutation relations
$\{\psi,\psi^\dagger\}=\epsilon=[a,a^\dagger]=-i[x,\dot x]$ (with
$a=(x+i\dot x)/\surd{2}$) {\it and} a vacuum\footnote{It has
recently been suggested that the non-perturbative definition of
quantum gravity in dS be reexamined~\cite{Witten:2001}; in
particular the definition of the vacuum requires careful
consideration. Our local quantum field theoretic computation
ignores such subtleties.} $\psi\,|0\rangle=0=a\,|0\rangle$,
positivity of norms requires $\epsilon>0$. For quantum field
theories, exactly the same criterion can be applied, but now in a
distributional sense.

The local canonical (anti)commutators for quantum fields with
$s\leq 2$ in cosmological spaces were presented long ago
\cite{Higuchi:1987py}. Let us summarize the relevant results:

For $s\geq1$, away from the gauge invariant boundaries, the
constraints~\eqn{const} and~\eqn{sticky} imply the
(gamma-)traceless--transverse conditions
 \be
D.\psi=0=\g.\psi\, ,\;\;\;  D.\phi_\n=0=\phi_\r{}^\r\, ,
\label{beauty}
 \ee
 which allow the field equations \eqn{RS}-\eqn{ES} to be rewritten as
\be (\frac{1}{2}\,\DF{3/2}+m)\,\psi_\m=0 \, ,\;\;\;
(\DB2-m^2+2\Lambda)\,\phi_{\m\n}=0\, . \label{captivate}
 \ee
  We must now write (anti)commutators for fields satisfying
both~\eqn{beauty} and~\eqn{captivate}. The former is easily
imposed using the (gamma-)\-traceless--transverse decompositions
\bea
\phi_\m^{\rm T}&=&\phi_\m-D_\m\,\frac{1}{D^2}\,D.\phi\, , \nn
\eea
 \be D.\phi^{\rm T}=0\, ; \label{kane} \ee \bea
\psi_\m^{\TT}&=&\psi_\m -\frac{1}{4}\,\g_\m\,\g.\psi
+D_{\wt\m}\,\frac{1}{3D^2+\L}\,(\gamma \cdot D\g.\psi-4D.\psi)\, ,\nn \eea
\be D.\psi^{\TT}=0=\g.\psi^{\TT}\, ; \label{T} \ee \bea
\phi^{\TT}_{\m\n}&=& \phi_{\m\n}
-D_{(\m}\,\frac{2}{D^2+\L}\,(D.\phi_{\n)})^{\rm T}
-\,\frac{1}{4}\,g_{\m\n}\,\phi_\r{}^\r\nn\\&&
-\,D_{\{\m}D_{\n\}}\,
\frac{4}{D^2(3D^2+4\L)}\,\Big[D.D.\phi-\frac{1}{4}\,D^2\phi_\r{}^\r\Big]\,
, \nn \eea \be D.\phi_\m^{\TT}=0=\phi^{\TT}_{\;\r}{}^\r\, ;
\label{abel} \ee
 where a tilde over an index denotes its
gamma-traceless part, {\it i.e.} $X_{\wt \m}\equiv
X_\m-\frac{1}{4}\,\g_\m \g.X$. and $\{\cdot\cdot\}$ denotes the
symmetric-traceless part of any symmetric tensor, {\it i.e.}
$X_{\{\m\n\}}\equiv
X_{(\m\n)}-\frac{1}{4}\,g_{\m\n}\,X_\r{}^\r$\,.

Therefore the (anti)commutators for spins $1\leq s\leq 2$ are
given by
 \bea
  \{\psi_\m(x),\ol\psi_\n(x')\}&\!\!=\!\!&
i\,\S3_{\m\n}(x,x';2m)-\frac{i}{4}\,\g_\m^x\,\S1(x,x';2m)\,\g_\n^{x'}
\nn \\&&
\qquad+\;\frac{i}{3m^2+\L}\,\D_\m^x\,\S1(x,x';2m)\,\overleftarrow\D_\n^{x'}
\, ,
\\
{}[\phi_{\m\n}(x),\phi_{\r\s}(x')]
&\!\!=\!\!&i\,\Dp2_{\m\n,\r\s}(x,x';m^2-2\L)
+\frac{2i}{m^2}\,D_\m^x\, D_\r^{x'}\, \Dp1_{\n\s}(x,x';m^2-2\L)\nn\\
&\!\!+\!\!\!&\frac{i}{m^2\,(3m^2-2\L)}\,\Big[2\,D^x_\m D_\n^x\,
D^{x'}_\r D_\s^{x'}+m^2(\L-m^2)\,g_{\m\n}^x\,g_{\r\s}^{x'}\nn\\&&
+\,m^2\,D^x_\m D_\n^x\,g_{\r\s}^{x'}+m^2\,g_{\m\n}^x\,D^{x'}_\r
D_\s^{x'} \Big]\;\Dp0(x,x';m^2-2\L)\, .\nn\\&& \label{twosies}
\eea
 For brevity, we have suppressed obvious symmetrizations
$(\m\n)$ and $(\r\s)$ on the right hand side of~\eqn{twosies}. The
field equations~\eqn{captivate} have been used throughout to
eliminate any factors $D^2$ appearing in the
(gamma-)traceless--transverse
decompositions~\eqn{kane}-\eqn{abel}, since the higher
distributions are also onshell
 \bea
(\DB{n}+m^2)\,\Dp{n}_{\m_1\ldots\m_n,\n_1\ldots\n_n}(x,x';m^2)&\equiv&0\, ,\\
(\DF{n+1/2}+m)\,
S^{(n+1/2)}_{\m_1\ldots\m_n,\n_1\ldots\n_n}(x,x';m)\;\;&\equiv&0\,
. \eea
 The identity
$D_{\wt\m}^x\,\S1(x,x';2m)=\D^x_\m\,\S1(x,x';2m)$ is also useful.
The (anti)commutators (3.6)-\eqn{twosies} are the difference
between advanced and retarded propagators. The distributions
$\Dp2$ and $\S3$ above satisfy
 \bea
\S3_{\m\n}(x,x';m)\,\overleftarrow\D^\n_{x'} & = &
-\D^x_\m\,\S1(x,x';m)\, ,\\
\S3_{\m\n}(x,x';m)\,\g^\n_{x'} & = & \g^x_\m\,\S1(x,x';m)\, ,\\
D^\r_{x'}\,\Dp2_{\m\n,\r\s}(x,x',m^2) & = &
-D^x_{(\m}\,\Dp1_{\n)\s}(x,x';m^2)\, ,\\
g^{\r\s}_{x'}\,\Dp2_{\m\n,\r\s}(x,x',m^2) & = &
g_{\m\n}^x\,\Dp0(x,x',m^2)\, ,
 \eea
with boundary conditions
 \be
  \S3_{\m\n}(x,x')\Big|_{x^0=x'{}^0}=
\frac{g_{\m\n}\,\g^0}{\sqrt{-g}} \,\delta^3(\vec x-\vec x\,')\; ,
\label{ping}
 \ee
$$
\frac{d}{dx'{}^0}\Dp2_{\m\n,\r\s}(x,x')\Big|_{x^0=x'{}^0}=
\frac{g_{(\m(\r}\,g_{\s)\n)}}{\sqrt{-g}} \,\delta^3(\vec x-\vec
x\,')\;,\quad\ \Dp2_{\m\n,\r\s}(x,x')\Big|_{x^0=x'{}^0}=0\, .
$$
Equipped with the above tools, one can easily uncover the
nonunitary regions. Starting with the (anti)commutators
(3.6)-\eqn{twosies}, the aim is to determine whether the
distributions on their right hand sides have definite sign in the
dangerous lower helicity sectors.

For concreteness we work in the simple synchronous dS metric
 \be
ds^2=-dt^2+e^{2Mt}\,d\vec{x}^2\, , \qquad M\equiv \sqrt{\L/3}\ ,
\label{easy} \ee
 and concentrate on the equal time
(anti)commutators of the time components of the fields (and their
time derivatives). While the metric~\eqn{easy} does not cover the
entire dS space, nor is it real when continued to negative AdS
values of $\L$,  these disadvantages are outweighed by its
simplicity. Selecting the lowest helicity components by looking at
time components of fields, a simple computation reveals that
 \bea
\{\psi_0(t,\vec x),\psi^\dagger_0(t,\vec x\,'\}&=&
-\,\frac{\nabla^2}{3m^2+\L}\;\frac{1}{\sqrt{-g}}\,\delta^3(\vec
x-\vec x\,')\;,
\\\nn \\
{}[\phi_{00}(t,\vec x),\dot\phi_{00}(t,\vec{x}\,')]&=&
\frac{2\,\nabla^4}{m^2\,(3m^2-2\L)}\;\frac{i}{\sqrt{-g}}\,\delta^3(\vec
x-\vec x\,')\;, \label{sung}
 \eea
  where $\nabla^2\equiv
g^{ij}\d_i\d_j =e^{-2Mt}\,\vec \d\,^2$ is a negative operator. The
final equation~\eqn{sung} agrees with the detailed massive $s=2$
Dirac analysis presented in \cite{Higuchi:1987py}. Our derivation
only requires writing out the Laplace and Dirac operators
explicitly in the metric~\eqn{easy} and using the field
equations~\eqn{captivate} to maximally eliminate time derivatives,
after which the boundary conditions \eqn{ping} may be applied.

The interpretation of equations (3.16)-\eqn{sung} is as follows:
Positivity of the distributions on the right hand sides is
completely determined by the respective denominators $3m^2+\L$ and
$3m^2-2\L$, precisely the factors appearing in the Bianchi
identities of the previous Section. For the various spins, we
learn:
\begin{itemize}
\item \underline{Spin~3/2:\ } The model is unitary in the region
$m^2>-\L/3$ which includes the Minkowski background. Strictly
massless, gauge invariant unitary models are found along the AdS
line $m^2=-\L/3$. The region $m^2<-\L/3$ is non-unitary. In
contrast to flat space, the $m^2=0$ theories are {\it massive}
when $\L\neq0$ and even non-unitary for negative (AdS) values of
$\L$.
\item \underline{Spin~2:\ } Models with $m^2>2\L/3$ are unitary.
There are now two lines of gauge invariant theories; the usual
linearized cosmological Einstein theory at $m^2=0$ and a partially
massless theory~\cite{Deser:1983tm} at $m^2=2\L/3$. Both are
unitary but the region $m^2<2\L/3$ is not.
\end{itemize}

Finally, as promised, we address the concern that, strictly, the
metric~\eqn{easy} applies only to dS. On the one hand, given that
(i) the final results are a function of the real variable $\L$
only and (ii) the picture presented here is backed up by the
emergence of Bianchi identities, there can be no doubt of its
correctness. However, for complete certainty, we repeat, as an
example, the $s=3/2$ computation in the metric
($M\equiv\sqrt{\L/3}$) \be
ds^2=-dt^2+\cosh^2(Mt)\,\Big\{dr^2+\frac{1}{M^2}\,\sin^2(Mr)\,(d\theta^2+
\sin^2\theta\,d\phi^2)\Big\}\, . \label{buoy} \ee Upon rescaling
$r\rightarrow\rho/M$, the three-metric
$d\Omega^2=d\r^2+\sin^2\r\,(d\theta^2+\sin^2\theta\,d\phi^2)$ is
seen to describe a unit three-sphere. We prefer the initial
parametrization, however, since for pure imaginary values  of $M$,
the cosmological constant $\L$ is  negative and the metric
continues to AdS. Performing a calculation analogous to the one
above we find
 \bea
\{\psi_0(t,r,\theta,\phi),\psi^\dagger_0(t,r',\theta',\phi')\}&=&
\nn\\&&\hspace{-4.2cm}
\frac{\cosh^{-2}(Mt)\,(-\,^{(3)}\!D^2-\L/4)}{3m^2+\L}
\;\frac{1}{\sqrt{-g}}\,\delta(r-r')
\,\delta(\theta-\theta')\,\delta(\phi-\phi')\;.\nn\\
\eea
 The operator ${}^{(3)}\!D^2$ is the square of the intrinsic
3-dimensional covariant derivative (Laplace--Beltrami operator)
acting on a spinor. In dS the operator $-\,^{(3)}\!D^2-\L/4$ is
not manifestly positive. However (in our parametrization) the
eigenvalues of ${}^{(3)}\!D^2$ acting on spinors are
$(\L/3)(-l(l+2)+1/2)$ with $l\geq 1/2$ (see, {\it
e.g.}~\cite{Higuchi:1987py}), and the highest eigenvalue is
precisely $-\L/4$. Hence the operator $-\,^{(3)}\!D^2-\L/4$ is
indeed positive and in dS we may draw precisely the conclusions
given above. Now, continuing the metric~\eqn{buoy} to AdS space
the same result holds for the local anticommutator except the
3-space is a hyperboloid. Nonetheless, (assuming we can neglect
spatial boundary terms), both $-{}^{(3)}\!D^2$ and $-\L/4$ are now
separately positive, and unitarity is determined by the sign of
the denominator $3\L+m^2$. This concludes our derivation of the
unitarily forbidden regions for spins $s=3/2,2$.

We emphasize that once one knows the gauge lines and their
corresponding Bianchi identities, our unitarity results in fact
follow by inspection: Whenever a coefficient in a massive
constraint vanishes and then becomes negative, all the
corresponding lower helicity modes are first excised by the
accompanying gauge invariance and thereafter reemerge with
opposite norms. Therefore, starting from the unitary Minkowski
region, it is easy to map out the unitarily allowed and forbidden
regions, as shown in the Figure. Furthermore, for higher spin
partially massless theories to be unitary, the ordering criterion
for the gauge lines, discussed in the introduction, must hold. A
simple example is provided by the $s=2$ strictly massless
($m^2=0$, linearized graviton) theory for $\L>0$:  To reach it
starting from the unitary Minkowski region, one must pass through
the unitarily forbidden region $0<m^2<2\L/3$. Nonetheless, the
theory is unitary, since the highest helicities $\pm2$ are left
untouched by the unitarity flip of the helicity $0$ mode across
the $m^2=2\L/3$ gauge line.

\section{\hspace{-.3cm}Partially$\!$ Massless$\!$
Spin$\!$~2: Canonical$\!$ Analysis}

\renewcommand{\theequation}{4.\arabic{equation}}
\setcounter{equation}{0}

\label{whale}

At the partially massless dS boundary, $m^2=2\L/3=2M^2$, we showed
that the scalar constraint~\eqn{sticky} is a Bianchi identity; as
such it removes the fifth DoF, leaving the 4 physical DoF
corresponding to helicities $\pm2,\pm1$. In this Section we prove
this claim via an explicit canonical analysis\footnote{A detailed
canonical analysis of massive $s=2$ for general $m^2$ is given in
the fourth reference in ~\cite{001}; an early attempt can be found
in~\cite{Bengtsson:1995vn}.}. Our method is similar to that
originally used to prove the stability of massless cosmological
gravity~\cite{Abbott:1982ff}.

A possible starting point is the second order massive $s=2$ action
in equation~\eqn{ramit}. Equivalently (and much simpler) one can
begin with the first order ADM form of cosmological Einstein
gravity, \bea S_{\L+E}&=&\int d^4x\; \Big[\;\pi^{ij}\,\dot g_{ij}
+N\,\sqrt{g}\;\Big({}^{(3)}\!R-2\L\Big)+2N_iD_j\pi^{ij}
\nn\\&&\qquad\qquad\quad +\frac{N}{\sqrt{g}}\,\pi^{ij}\pi^{lm}\,
\Big(\frac{1}{2}\,g_{ij}g_{lm}-g_{il}g_{jm}\Big) \Big]\,
,\label{einst} \eea then linearize around a dS background and add
by hand an explicit mass term. Here $g$ is the determinant of the
3-metric $g_{ij}$ and $N\equiv(-g^{00})^{-1/2}$, $N_{0i}\equiv
g_{0i}$. We take the synchronous dS metric~\eqn{easy}, denoted
$ds^2=-dt^2+\ol g_{ij}dx^i dx^j$ in this Section, reserving
$g_{ij}$ for the dynamical 3-metric which we linearize as \be
g_{ij}\equiv\ol g_{ij}+\phi_{ij}\, ,\qquad\ol g_{ij}\equiv
f^2(t)\,\delta_{ij} \; ,\qquad f(t)\equiv e^{Mt}\, . \ee The
remaining fields are linearized as \be
\pi^{ij}\equiv\ol\pi^{ij}+P^{ij}\;,\quad\ol
\pi^{ij}\equiv-2Mf\,\delta^{ij}\;,\qquad N\equiv1+\wt n\, . \ee
(The background metric is block diagonal so no expansion is needed
for $N_i$.) In terms of these deviations, the mass term is \be
S_m=-\frac{m^2}{4}\,\int d^4x\,\sqrt{\ol g}\,
\Big(\phi_{\m\n}\phi_{\r\s}\ol g^{\m\r} \ol
g^{\n\s}-(\phi_{\m\n}\ol g^{\m\n})^2\Big)\, ,\label{mass} \ee here
$\phi_{0i}\equiv N_i$, $\ol g\equiv\det \ol g_{ij}=-\det\ol
g_{\m\n}$ and $\phi_{00}\equiv g_{00}+\ol g_{00}=-(1+N^2)$. The
final action is the sum $S=S_{\L+E}+S_m$, discarding any terms of
higher than quadratic order in (dynamical) fields.

Notice that the only explicit time dependence of the integrand
of~\eqn{mass} is through $f^{-1}$. Indeed it proves useful to make
the field redefinition \bea \phi_{ij}\equiv f^{1/2}\,h_{ij}\;
,\qquad P^{ij}\equiv f^{-1/2}\,p^{ij}\, . \nn \eea \be N_i\equiv
f^{1/2}\,n_i\;,\qquad \wt n=f^{-3/2}\,n\, . \label{redef} \ee The
cost is an extra contribution generated by the symplectic term
of~\eqn{einst}, $P^{ij}\dot\phi_{ij}$ $\rightarrow$ $p^{ij}\dot
h_{ij} + (M/2) p^{ij}h_{ij}$. A dividend is that the only explicit
time dependence in what follows will be through $\nabla^2\equiv
\ol g^{ij}\d_i\d_j\equiv f^{-2}\nabla_0^2$. Index contractions are
just with $\delta_{ij}$, all quantities are now in $(3+1)$ form.

Next examine the mass term \be S_m=\int d^4x\,\Big[
-\frac{m^2}{4}\,(h_{ij}^2-h_{ii}^2)+\frac{m^2}{2}\,N_i^2+m^2
\,n\,h_{ii}\Big]\ . \ee Were it not for the term proportional to
$N_i^2$, the field $N_i$ would be a  Lagrange multiplier for 3
constraints (as is the case for the $m=0$ strictly massless
theory). Instead, when $m\neq0$ we must   integrate out $N_i$ via
its algebraic equation of motion. The field $n$, however, only
appears linearly and remains a Lagrange multiplier for the
constraint \be \Big[\sqrt g\,({}^{(3)}\!R-2\L)+m^2\,h_{ii}
+\frac{1}{\sqrt{g}}\,\pi^{ij}\pi^{lm}\,
(\frac{1}{2}\,g_{ij}g_{lm}-g_{il}g_{jm})\Big]_{\rm \sss
linearized}=0\, . \label{connie} \ee For generic values of
$(m^2,\Lambda)$, this constraint eliminates one degree of freedom
from the 6 pairs $(p^{ij},h_{ij})$ leaving 5 physical helicities
$(\pm2,\pm1,0)$. Our aim now is to show that a (single) further
constraint emerges at the gauge invariant value $m^2=2M^2$.

We decompose the fields $h_{ij}$ and $p_{ij}$ according to their
helicity and drop (for now) the helicity $\pm2$
traceless-transverse $(h_{ij}^{tt},p_{ij}^{tt})$ {\it and}
helicity $\pm1$ transverse $(h_{i}^{t},p_{i}^{t})$ modes since
they manifestly decouple from the 2 remaining helicity 0 modes (to
the quadratic order used here). The latter are defined by the
projection \be
h_{ij}=\frac{1}{2}\,\Big(\delta_{ij}-\frac{\d_i\d_j}{\nabla_0^2}\Big)\,h_T
+\frac{\d_i\d_j}{\nabla_0^2}\,h_L\, , \ee and similarly for
$p_{ij}$. (Note that under the integral $\int A_{ij}\,B_{ij}=
\frac{1}{2}\,\int A_T B_T+\int A_L\,B_L$.) In terms of these
variables the linearized constraint is \be {\cal
C}\equiv\nabla^2h_T+2M\,(p_T+p_L)-(m^2-2M^2)\,(h_T+h_L)=0\,
.\label{twang} \ee Note that the leading term comes from the
linearized 3-dimensional curvature scalar and that both the
cosmological $-2\L\sqrt{g}$ and momentum squared terms
in~\eqn{connie} contribute to the final term in~\eqn{twang}, which
vanishes on the critical gauge line $m^2=2M^2$. Henceforth we
concentrate on the critical case and eliminate $m^2$ via this
relation.

Next we write out the quadratic action $S=S_{E+\L}+S_m$
remembering the constraint~\eqn{twang} which we solve as \be
p_T=-p_L-\frac{\nabla^2}{2M}\;h_T\, . \ee Observe that the
symplectic terms become \be p_{ij}\,\dot
h_{ij}=\frac{1}{2}\,p_T\dot h_T+ p_L\dot h_L= p_L\dot q-
\frac{1}{4}\,h_T\,\nabla^2\,h_T\; ,\qquad q\equiv
h_L-\frac{1}{2}\,h_T\, . \ee (suppressing integrations
throughout). Upon eliminating $h_L=q+h_T/2$ in favor of $q$ and
$h_T$, the action depends only on the 3 variables $(p_L, q,h_T)$
and its most general form is \be S(p_L, q,h_T)-  p_L\,\dot q=
\frac{1}{2}\,A\, h_T^2\,+\,B\,h_T\,+\,C\ , \ee where $A$ is
constant, $B$ linear and $C$ quadratic in $(p_L,q)$. If $A=0$, we
have an additional constraint $B(p_L,q)=0$ and no zero helicity
DoF remain, whereas for non-zero $A$, $h_L$ can be removed via its
algebraic field equation leaving behind one zero helicity DoF. In
fact, $A$ does vanish on the critical line $m^2=2\L/3$ so this
model describes helicities $(\pm2,\pm1)$ only. Indeed, a lengthy
calculation yields \be S(p_L, q,h_T)-  p_L\,\dot q=
\frac{1}{2}\,\Big(\frac{p_L}{M}-q\Big)\nabla^2\,h_T
\,+\,\Big(\frac{p_L}{M}-q\Big)
\Big[\!\Big(\frac{\nabla^2}{m^2}-\frac{3}{2}\Big)\,p_L
-\Big(\!\nabla^2-m^2\Big)\,q\Big]\ . \ee As claimed, $A=0$ and
even the zero helicity Hamiltonian vanishes once the Lagrange
multiplier $h_T$ is integrated out.

Let us now examine the remaining helicities $(\pm2,\pm1)$. A
series of canonical transformations yields a simple action \be
S_{(\pm2,\pm1)}=\sum_{\ve=(\pm2,\pm1)}\,\Big\{ p_{\ve}\,\dot q_\ve
- \frac{1}{2}\, \Big[ \,p_\ve^2\ + \
q_\ve\,\Big(\!-\nabla^2-\frac{M^2}{4}\Big)\,q_\ve \ \Big]\Big\}\,
. \label{hell} \ee Notice again, all time dependence is through
$\nabla^2$ in the Hamiltonian. The field equations are \be
p_\ve=\dot q_\ve\ , \qquad
\Big(-\frac{d^2}{dt^2}+\nabla^2-\frac{M^2}{4}\ \Big)\,q_\ve=0\ .
\label{picky} \ee The covariant field equation~\eqn{captivate}
evaluated at $m^2=2M^2$ is $(D^2-4M^2)\,\phi_{\m\n}=0$. Consider,
for example, helicities $\pm 2$, for which
$\d^i\phi_{ij}=0=\phi_i{}^i$. In this frame the
transverse-traceless part of the covariant field equation reads
\be \Big(-\frac{d^2}{dt^2}\ + \ M\,\frac{d}{dt}\ +\ \nabla^2\
\Big)\,\phi_{ij}=0\, .\label{icky} \ee The action~\eqn{hell} was
obtained by the same rescaling as in~\eqn{redef}, namely,
$\phi_{ij}=f^{1/2}\,q_{ij}$. The factor $f^{1/2}$ is precisely the
integrating factor which removes the single time derivative from
equation~\eqn{icky} at the cost of a term $-M^2/4$, {\it i.e.}
equations~\eqn{icky} and~\eqn{picky} are identical (helicities
$\pm1$ agree via a similar calculation).

Stability of the partially massless theory requires that it
possess a conserved, positive, energy function. The latter can be
obtained by an argument similar to that  given
in~\cite{Abbott:1982ff} for the strictly massless $s=2$ theory:
The Hamiltonian in~\eqn{hell} is not conserved because of the
explicit time dependence of $\nabla^2$. However, inside the
intrinsic dS horizon at $(fMx^i)^2=1$, the background
metric~\eqn{easy} possesses a timelike Killing vector \be
\xi^\m=(-1,Mx^i) \Longrightarrow \xi^2=-1+(fMx^i)^2\ . \ee
Therefore, the energy associated with time evolution in this
Killing direction \be E=T^0{}_\m\xi^\m= H-Mx^i\,\Big[p_\ve\,\d_i
q_\ve-\frac{1}{2}\,\d_i\,(p_\ve q_\ve)\Big] \ee satisfies $\dot
E=0$ ($H$ is the Hamiltonian in~\eqn{hell} and we have suppressed
the sum over helicities $\ve$). Furthermore, writing out $E$
explicitly and relabeling the variable $p_\ve\rightarrow
p_\ve+(3M/2)q_\ve$ gives \be E=\frac{1}{2}\,\Big(\wh
x^i\,p_\ve\Big)^2 +\frac{1}{2}\,\Big(f^{-1}\d_i\,q_\ve\Big)^2 -f M
|x|\,\Big(\wh x^i\,p_\ve\Big)\,\Big(f^{-1}\d_i\,q_\ve\Big)
+\frac{1}{2}\,(2M^2)\,q_\ve^2\, , \ee with $x^i\equiv |x|\,\wh
x^i$. The last (mass) term is manifestly positive and the first
three terms are positive by the triangle equality whenever \be
f\,M \,|x|<1\ , \ee that is, inside the physically accessible
region.

A final interesting feature of the partially massless $s=2$ theory
is null propagation. The dS metric is conformally flat and it can
be shown that the $m^2=2\L/3$ theory propagates on its null
cone~\cite{Deser:1983tm}. We will later see that this property is
shared by all partially massive systems of higher spins.

\section{Higher Spins}

\renewcommand{\theequation}{5.\arabic{equation}}
\setcounter{equation}{0}

\label{dolphin}

Having seen that the $s=2$ field equation ${\cal G}_{\m\n}$ yields
both single divergence, $D.{\cal G}_{\n}$, and double divergence,
$D.D.{\cal G}$, Bianchi identities we are led to inquire whether
even higher divergence Bianchi identities occur for $s>2$. The
answer (see the third paper of \cite{001}) is yes: In addition to
the usual massive and strictly massless possibilities, a spin $s$
field in (A)dS can be partially massless with propagating
helicities $(\pm s,\pm(s-1),\ldots,\pm(s-t))$ ($t<s$). In this
section, we concentrate on the explicit analysis of $s = 5/2$ and
3 as examples of the generic cases.

\vspace{.4cm}

\noindent{\large\bf Spin~5/2}

\vspace{.4cm}

The $s=5/2$ spinorial field equation has two open indices, so as
for $s=2$, there are two possible Bianchi identities; they appear
along the AdS gauge lines $m^2=-4\L/3$ and $m^2=-\L/3$. The former
is the strictly massless theory with helicities $\pm5/2$ whereas
the novel gauge invariance of the latter removes only the lowest
$\pm1/2$ leaving helicities $(\pm5/2,\pm3/2)$. Since the massless
gauge lines all lie in AdS (just as for their $s=3/2$
counterpart), the $(m^2,\L)$ half-plane is divided into 3 regions.
Only the $m^2>-4\L/3$ one including Minkowski space, is unitary.

The $s=5/2$ action and field equations are
 \be {\cal
L}=-\,\sqrt{-g}\;\ol\psi^{\m\n}\,{\cal R}_{\m\n}
-\,\sqrt{-g}\;\ol\chi\,{\cal R}_5\, , \label{rosella} \ee \bea
{\cal R}_{\m\n}&=&(\DF{5/2}-2\gamma \cdot
D)\,\psi_{\m\n}+g_{\m\n}\,\g.D.\psi
+(D_{(\m}\g_{\n)}-\frac{1}{2}\,g_{\m\n}\,\gamma \cdot D)\,\psi_\r{}^\r\nn\\
&+&m\,(\psi_{\m\n}-2\,\g_{(\m}\g.\psi_{\n)}-\frac{1}{2}\,g_{\m\n}\,
\psi_\r{}^\r) -\frac{5}{12}\,\mu\,g_{\m\n}\,\chi=0\, ,
\label{RR}\\
{\cal R}_5&=&-\a\,(\gamma \cdot
D-3m)\,\chi-\frac{5}{12}\,\mu\,\psi_\r{}^\r=0\, . \label{R5}
 \eea
Minimal coupling alone does not provide equations of motion
describing the $6=2s+1$ massive DoF: An additional non-minimal
coupling contained by the term $\mu \,\ol \chi\, \psi_\r{}^\r$ is
necessary. In fact, to achieve a proper set of constraints
requires fixing the auxiliary coupling to
 \be
\mu^2=\frac{12\a}{5}\,(m^2+4\L/3)\ . \label{mu}
 \ee
  Here we
encounter a new subtlety. Implicitly we have so far assumed that
physical models live in the half plane $m^2>0$, since for fermions
negative $m^2$ implies a non-hermitean mass term, and for bosons,
one that is unbounded below. While there are regions with both
$m^2<0$ and the correct sign for anticommmutators, the dynamics is
non-unitary there. Therefore we continue to require $m^2>0$ and
examine the relation~\eqn{mu}. Since hermiticity of the
action~\eqn{rosella} demands $\mu^2>0$, we find two regions: (i)
$m^2>-4\L/3>0$, (ii) $0<m^2<-4\L/3$. Up until now, $\a$ was a free
parameter which we could set to $\pm1$. In the region (i),
$m^2+4\L/3>0$ so we must take $\a=+1$. In region (ii), hermiticity
of the action can be maintained at the cost of changing the sign
of $\a$ to  $\a=-1$ (the actions are then different in each
region).  In either case, we will find that region (ii) is
unitarily forbidden.

Before continuing, it is interesting to compare these difficulties
to $s=3/2$ and the problem of constructing dS supergravities. As
we have shown, $s=3/2$ is unitary for $m^2\geq -\L/3\geq0$ and the
boundary is the strictly massless AdS theory corresponding to
cosmological supergravity. As one follows the massive theory into
dS, the canonical anticommutators  all have the correct sign for
unitary representations. In fact, keeping $\L>0$, there is no
obstruction at the level of anticommutators to continuing to
$m^2<0$ until the branch of the line $m^2=-\L/3$ with $m^2<0<\L$
is reached. The theory there is formally supersymmetric ({\it
i.e.}, strictly massless) but the action is no longer hermitean,
which is an example of the general statement that dS
supergravities do not exist~\cite{Pilch:1985aw}. One might
speculate that this clash is generic to higher spin fermions.

Returning to the $s=5/2$ field equations, for generic $(m^2,\L)$
the constraints \bea {\cal C}_{\wt \n}&\equiv& D.{\cal R}_{\wt
\n}+\frac{1}{4}\,m\,\g.{\cal R}_{\wt \n} \label{C51}
\\
&=& -\frac{5}{4}\,(m^2+4\L/3)\,\g.\psi_{\wt \n}
-\frac{5}{12}\,\mu\,D_{\wt \n}\,\chi\, ,\label{coleslaw}\\
{\cal C}&\equiv& D.{\cal C}+ \frac{5}{16}\,(m^2+4\L/3)\,{\cal
R}_\r{}^\r -\frac{5}{16}\,\a\mu\,(\gamma \cdot D+3m)\,{\cal R}_5
\label{c1}\\&=& -\frac{10}{3}\,\mu\,(m^2+\L/3)\,\chi \
,\label{C52} \eea ensure that the model describes $6=2s+1$ massive
DoF. Along the AdS lines \be m^2=-4\L/3\, \qquad m^2=-\L/3 \ee the
constraints~\eqn{coleslaw} and~\eqn{C52} transmute to Bianchi
identities with respective (distinct) gauge invariances
 \bea
\delta\psi_{\m\n}=D_{(\m}\varepsilon_{\wt
\n)}+\frac{1}{2}\,\sqrt{\frac{-\!\L\,}{3}}
\,\g_{(\m}\,\varepsilon_{\wt \n)}\, ,&\delta\chi=0\, ,
\label{ar}\\
\delta\psi_{\m\n}=D_{(\m}D_{\wt \n)}\,\varepsilon
+\frac{5\L}{16}\,g_{\m\n}\,\varepsilon\, ,\qquad&
\delta\chi=-\frac{1}{8\a}\,\sqrt{15\a\L}\,(\gamma \cdot D+\sqrt{-3\L})
\,\varepsilon \, .\qquad\label{ra} \eea
 The vector-spinor Bianchi
identity~\eqn{C51} at $m^2=-4\L/3$ implies strict masslessness
(propagating helicities $\pm5/2$) since its invariance removes
helicities $(\pm3/2$, $\pm1/2)$. Notice also that $\mu=0$ on the
strictly massless line so, as claimed above, the spinor auxiliary
$\chi$ decouples there. The novel spinor Bianchi
identity~\eqn{C52} at $m^2=-\L/3$ and invariance~\eqn{ra} removes
helicities $\pm1/2$ leaving a partially massless theory of
helicities $(\pm5/2,\pm3/2)$.

Once again, the coefficients $(m^2+4\L/3)$ and $(m^2+\L/3)$
appearing in the constraints~\eqn{C51} and~\eqn{C52} control the
positivity of equal time anticommutators. Therefore, since the
gauge lines all lie in AdS, the $(m^2,\L)$-plane is divided into 3
regions; only the one including Minkowski space $m^2>-4\L/3$ is
unitary. Although the strictly massless, AdS, $m^2=-4\L/3$ theory
is unitary, the partially massless one is not, as it fails the
line ordering requirement: Starting from the unitary Minkowski
region where all norms are positive, one would like first to
traverse the line $m^2=-\L/3$, but that is only possible in dS
with negative $m^2$ (imaginary values of $m$ violate hermiticity
of the action and unitary evolution). Crossing the AdS strictly
massless line $m^2=-4\L/3$ first flips the norm of both lower
helicities $(\pm3/2,\pm 1/2)$ so the partially massless AdS theory
cannot be unitary. Ironically, were negative values of $m^2$ not
prohibited, we could traverse the lines in the correct order in
dS. This observation lends weight to the speculation that the
unitarity difficulties of partially massless theories are peculiar
to half integer spins. Indeed, in the next Section we exhibit the
bosonic $s=3$ example, which enjoys two partially massless unitary
dS lines.

\vspace{.4cm}

\noindent{\large\bf Spin~3}

 \vspace{.4cm}

Spin~3 is the first example of a system with two new Bianchi
identities over and above the usual one at $m^2=0$. The action and
field equations are
 \be {\cal
L}=\,\frac{1}{2}\,\sqrt{-g}\,\phi^{\m\n\r}\,{\cal G}_{\m\n\r}
\,-\,\frac{3}{8}\,\sqrt{-g}\,\chi\,{\cal G}_{5}\ , \ee \bea {\cal
G}_{\m\n\r}&=&(\DB3-m^2+16\L/3)\,\phi_{\m\n\r}-3D_{(\m}D.\phi_{\n\r)}+
3D_{(\m}D_{\n}\phi_{\r)\s}{}^\s\nn\\
&-&3g_{(\m\n}\,\Big( (\DB1-m^2+11\L/3)\,\phi_{\r)\s}{}^\s
-D.D.\phi_{\r)\s}{}^\s
+\frac{1}{2}\,D_{\r)} D.\phi_\s{}^\s\Big)\,\nn\\
&+&\,\frac{3m}{4}\,g_{(\m\n}\,D_{\r)}\chi= 0\,  ,\label{w}\\
{\cal
G}_5&=&\frac{3}{2}\,(\DB0-4m^2+8\L)\,\chi+m\,D.\phi_\s{}^\s\, =
0\, . \label{izard} \eea
 The field $\phi_{\m\n\r}$ is a symmetric
3-tensor and the auxiliary field $\chi$ decouples at $m=0$ (the
strictly massless theory). Fixing the ordering of covariant
derivatives as shown and requiring constraints to remove all but
the physical $7=2s+1$ degrees of freedom, uniquely specifies all
terms with an explicit $\L$-dependence. Indeed, we find the
following constraints
 \bea {\cal B}_{\{\n\r\}}
\!\!\!\!&\equiv&D.{\cal G}_{\{\n\r\}}=-\frac{1}{2}\,m\,
\Big(\,D_{\{\n}D_{\r\}}\chi+2m\,D.\phi_{\{\n\r\}}
-4\,m\,D_{\{\n}\phi_{\r\}\s}{}^\s\Big)\, ,%\nn\\
\label{B1} \\
{\cal B}_\r\!\!&\equiv& D.{\cal B}_\r-\frac{m}{4}\,D_\r{\cal G}_5
+\frac{m^4}{4}\,{\cal G}_{\r\s}{}^\s=
\frac{5}{8}\,m\,(3m^2-4\L)\,(D_\r\chi+\frac{2}{3}\,m\,\phi_{\r\s}{}^\s)
\, ,
\nn\\ \\
{\cal B}\!\!&\equiv& D.{\cal B}-\frac{5}{12}\,m\,(3m^2-4\L)\,{\cal
G}_5=
\frac{5}{2}\,m\,(3m^2-4\L)\,(m^2-2\L)\,\chi\, .%\nn \\
\label{B3} \eea
 The explicit tensorial structures on the right
hand sides of~\eqn{B1}-\eqn{B3} exhibit the successive splitting
of the prefactors to $m$, namely $(3m^2-4\L)$ and $(m^2-2\L)$,
thanks to the additional parameter $\L$. Therefore, in addition to
the usual massless theory at $m=0$ there are new gauge invariant
systems at $m^2=4\L/3$ and $m^2=2\L$, since whenever these
prefactors vanish, the corresponding constraints
in~\eqn{B1}-\eqn{B3} become Bianchi identities with accompanying,
respective,  gauge invariances
 \bea
  \delta
\phi_{\m\n\r}=D_{(\m}\,\xi_{\{\n\r\})} , && \delta\chi=0\; ;
\label{holeyghost}\\
\delta \phi_{\m\n\r}=D_{(\m}D_{\{\n}\,
\xi_{\r\})}+\frac{\L}{3}\,g_{(\m\n}\,\xi_{\r)} , && \delta
\chi=-\frac{2}{3}\sqrt{\frac{\L}{3}}\,D.\xi
\; ;\label{sun}\\
\delta \phi_{\m\n\r}=D_{(\m}D_{\{\n}D_{\r\})}\, \xi
+\frac{\L}{2}\,g_{(\m\n}D_{\r)}\,\xi , && \delta
\chi=-\frac{2}{3}\sqrt{\frac{\L}{2}}(D^2+\frac{10\L}{3})\xi
.\label{fader}\quad \eea
 The new gauge invariant lines bound
regions in the $(m^2,\L)$ half-plane whose unitarity properties
are determined by the signs of the prefactors $(3m^2-4\L)$ and
$(m^2-2\L)$. To analyze these new properties, decompose the
$7=2s+1$ physical DoF into helicities $(\pm3,\pm2,\pm1,0)$. We
find the following ``phase'' structure of the $(m^2,\L)$
half-plane
\begin{itemize}
\item \underline{$m^2>2\L>0$} : This region includes Minkowski
space and is clearly unitary. All helicities $(\pm3,\pm2,\pm1,0)$,
propagate with positive norm.
\item \underline{$m^2=2\L$} : A partially massless theory
appears since the scalar constraint ${\cal B}=0$ is now a Bianchi
identity whose associated gauge invariance removes the scalar
helicity $0$ excitation. The remaining 6 DoF, $(\pm3,\pm2,\pm1)$,
propagate with positive norm since they are unaffected by the
scalar gauge invariance.
\item \underline{$4\L/3<m^2<2\L$} : Although all 7 DoF are now
again propagating, the scalar helicity $0$ mode reemerges {}from
the gauge boundary $m^2=2\L$ with negative norm (since the factor
$(3m^2-4\L)\,(m^2-2\L)$ appears in canonical commutators as a
negative denominator). This is a unitarily forbidden region.
\item \underline{$m^2=4\L/3$} : This partially massless theory has
Bianchi identities ${\cal B}=0={\cal B}_\r$ whose gauge
invariances excise the helicities $(\pm1,0)$. The remaining 4 DoF,
helicities $(\pm3,\pm2)$ propagate with positive norm.
\item \underline{$0<m^2<4\L/3$} : Again all 7 DoF propagate but
now the scalar helicity has again positive norm since its
denominator $(3m^2-4\L)\,(m^2-2\L)$ is again positive. However,
the region is still unitarily forbidden because now the vector
helicities $\pm1$ suffer a negative denominator $(3m^2-4\L)$.
\item \underline{$m^2=0$} : This is the unitary strictly massless model
with tensor Bianchi identity ${\cal B}_{\{\n\r\}}=0={\cal
B}_\r={\cal B}$. Only the uppermost helicities $\pm3$ remain. An
added subtlety is the remnant decoupled  auxiliary field $\chi$.
\end{itemize}
Notice how the uppermost helicity $\pm3$ always emerges unscathed
as a pair of positive norm states but, unlike Minkowski space,
splittings into theories with intermediate lower helicities,
rather than only the full complement of $2s+1$ states of the
massive theory, are possible. Furthermore, the ordering of gauge
lines is the same as for the $s=5/2$ example. But, unlike for
$s=5/2$, the lines can be traversed in the order required for
unitarity of the partially massless dS theories without recourse
to unphysical, negative, values of $m^2$.

The general pattern for $s>3$ should be clear; although the
complexity of auxiliary fields will make the details arduous to
follow, the counting is not.

%following from nullprop.tex

\renewcommand{\theequation}{6.\arabic{equation}}
\setcounter{equation}{0}

\section{{\hspace{-.2in}}Null$\!$ Propagation$\!$ of$\!$
 Partially$\!$ Massive$\!$ Systems}

 Here we show that the partial massless ({\it i.e.}, gauge
invariant) truncated multiplets propagate on the null cone, and
give a general spin formula for the $m^2:\L$ tunings that govern
these systems.

\vspace{.4cm}

\noindent {\large\bf Bosons}

\vspace{.4cm}

\label{bose}

Our method is direct: We solve the helicity $\pm s$ field
equations for all higher spins in the dS background (3.15). When
they are present in the spectrum, all lower helicities propagate
in exactly the same manner, so we need not treat them separately.
In our frame spatial slices are flat, which allows the usual
definition of helicity. Solutions to the field equations are of
Bessel type, and take the form
 \be \Big(\mbox{slowly
varying}\Big)\,\times\, \exp(i\w u+i\vec k\cdot\vec x)\, ,\qquad
\w^2=\vec k\,^2\, , \label{wave} \ee
 whenever the index $\nu$ of
the Bessel function is half integer. Here
 \be u\equiv
-\frac{f^{-1}(t)}{M} \ee
is the conformal time coordinate in terms
of which
 \be ds^2=\frac{1}{M^2u^2}\,\Big(-du^2+d\vec x\,^2\Big)\,
. \ee
 Therefore we can read off the theories propagating on the
null cone directly from the index $\nu$. We will explicitly find
null propagation for all the $s\leq3$ partially massless theories
discussed previously.

The onshell conditions for a massive spin $s$ field in (A)dS are
\be \Big(D^2-m^2-(2+2s-s^2)\,M^2\Big)\,\phi_{\m_1\ldots\m_s}=0\, ,
\qquad D.\phi_{\m_2\ldots\m_s}=0=\phi^\r{}_{\r\m_3\ldots\m_s}\, .
\label{bshell} \ee The parameter $m^2$ has been chosen so that
$m^2=0$ corresponds to the strictly massless (i.e., helicity $\pm
s$ only) theory. [For $s=0$, equation~\eqn{bshell} describes a
conformally improved scalar.]

The traceless-transverse part of a spatial tensor is denoted by
the superscript $tt$. We project out helicity $\pm s$ by computing
only the tt part of~\eqn{bshell}, which in the frame (3.15) reads,
 \be \Big(-\frac{d^2}{dt^2}+ (2s-3)\, M\,\frac{d}{dt}
+f^{-2}\, \vec \d\,^2 - m^2 +2(s-1)\,M^2
\Big)\,\phi^{tt}_{i_1\ldots i_s}=0\, . \ee Fourier transforming
$\vec \d\rightarrow i\vec k$, changing coordinates \be
z\equiv-\frac{kf^{-1}(t)}{M}=ku\, ,\qquad (k\equiv |\vec k|)
\label{change} \ee
 and the field redefinition (suppressing indices)
 \be \phi^{tt}\equiv z^{3/2-s} q\, , \ee yield Bessel's
equation \be \frac{d^2q}{dz^2}+\frac{1}{z}\,\frac{dq}{dz}
+\Big(1-\frac{\nu^2}{z^2}\Big)\,q=0\, , \label{Bessel} \ee
 with
index
 \be \nu^2=\frac{1}{4}+s(s-1)-\frac{m^2}{M^2}\, . \ee
 We may now read off the null propagating theories.

%\vspace{.4cm}

\newpage

\noindent {\bf Examples:}

\begin{enumerate}
\item {\em Conformal Scalar}: At $s=0=m$ we obtain $\nu=1/2$
and $q(z)=z^{-1/2}\,\exp(iz)$ which implies a solution of the
form~\eqn{wave}. The value $\nu=1/2$ also characterizes all the
higher spin ``conformal'' theories.
\item {\em Maxwell}: In $d=4$ the $s=1$, $m=0$ vector theory is conformal
and here $\nu=1/2$.
\item {\em Spin 2}: Spin~2 can be either strictly massless at $m=0$ or
partially massless when $m^2=2M^2$. The latter model, with its
accompanying scalar gauge invariance, takes the conformal value
$\nu=1/2$. Of course, the $m^2=0$ linearized cosmological Einstein
theory also propagates on the cone, but this is achieved by the
solution $\nu=3/2$ for which $q(z)=(z+i)\,z^{-3/2}\,\exp(iz)$.
\item {\em Spin 3}: Here there are 3 possibilities; the strictly massless
theory at $m=0$ for which $\nu=5/2$ ($\Rightarrow
q(z)=(z^2+3iz-3)\,z^{-5/2}\,\exp(iz)$), a partially massless one
with helicities $(\pm3,\pm2)$ at $m^2=4M^2$ with $\nu=3/2$ and
finally, a theory with a scalar gauge invariance and helicities
$(\pm3,\pm2,\pm1)$ with $m^2=6M^2$ and the conformal value
$\nu=1/2$. Clearly all these theories have null propagation.
\end{enumerate}
Noting that the value of $\nu$ can be associated with the type of
gauge invariance (for example, the conformal value $\nu=1/2$
always belongs to the scalar invariance) makes it likely that all
partially massless higher spin bosons propagate on the null cone.
The spin $s$ theory with helicities $(\pm n,\ldots, 0)$ removed
appears when \be m^2=M^2\,\Big(s(s-1)-n(n+1)\Big)\, , \ee and has
Bessel index $\nu=n+1/2$. All these gauge theories are unitary,
and (6.11) has indeed recently proved in \cite{russian}.

\vspace{.4cm}

\noindent {\large\bf Fermions}

\vspace{.4cm}

Partially massless fermionic theories are found in AdS. However,
we continue to work in dS because of the simplicity of the metric
(3.15). The results for the partially massless lines depend on
$m^2$ and $\L\equiv3M^2$ only and continue to AdS. Just as for the
lowest ``multiline'' $s=5/2$ case, only the strictly massless
helicity $\pm s$ gauge theory is unitary in AdS.

A massive spin $s\equiv \s+1/2$ fermionic field satisfies the
onshell conditions
 \be (\gamma \cdot D+m)\,\psi_{\m_1\ldots\m_\s}=0\,
\qquad D.\psi_{\m_2\ldots\m_\s}=0\ =\g.\psi_{\m_2\ldots\m_\s}\, .
\ee We choose the local Lorentz gauge \be e_0{}^{\underline 0}=1\,
, \qquad e_i{}^{\underline j}=f(t)\,\delta_i{}^{\underline j}\, ,
\ee
where underlined indices are flattened. The Dirac equation for
the (spatially transverse, gamma-traceless ``$tt$'') helicities
$\pm s$ reads \be \Big(
\frac{d}{dt}+(2-s)\,M-f^{-1}\,\g^{\underline{0j}}\,\d_j-\g^{\underline
0}\, m \Big)\,\psi^{tt}_{i_1\ldots i_\s}=0\, . \ee In the usual
large/small component basis (suppressing indices again)
 \be
\g^{\ul 0}=\left(
\begin{array}{rr}
-i&0\\ \;0&i
\end{array}\right)\, ,\qquad
\g^{\ul j}= \left(
\begin{array}{cc}
0&i\s^j \\ -i\s^j&0
\end{array}\right)\, ,\qquad
\psi^{tt}=\left(
\begin{array}{c}
\chi\\ \phi
\end{array}
\right)\, , \ee
 we can eliminate the small component $\phi$ and
obtain the second order equation for $\chi$ \be \Big(
-\frac{d^2}{dt^2}+(2s-5)\,M\,\frac{d}{dt}
+f^{-2}\,\vec\d\,^2-(s-2)(s-3)\,M^2-imM-m^2 \Big)\,\chi=0\, . \ee
The coordinate transformation~\eqn{change} and field redefinition
\be \chi\equiv z^{5/2-s}\,q\, , \ee yield Bessel's
equation~\eqn{Bessel} with index \be
\nu^2=\frac{1}{4}-\frac{im}{M}-\frac{m^2}{M^2}=
\Big(\frac{1}{2}-\frac{im}{M}\Big)^2\, . \ee Note that $m$ itself
will be imaginary for the partially massless lines when they are
in dS, so the appearance of an explicit $im$ here is appropriate.

\vspace{.4cm}

\noindent {\bf Examples:}

\begin{enumerate}
\item {\em Spin 1/2}: The $m=0$ spin~1/2 theory is well known to be Weyl
invariant, and indeed we find $\nu=1/2$, the conformal value.
\item {\em Spin 3/2}: As follows from linearizing cosmological supergravity,
spin~3/2 is strictly massless at $m^2=-M^2$. The choice of branch
$m=iM$, justified by the results, yields $\nu=3/2$. This is the
same value we found above for its strictly massless spin~2
superpartners.
\item{\em Spin 5/2}: The strictly massless theory is at $m^2=-4M^2$;
the choice $m=2iM$ yields $\nu=5/2$ and null propagation. The
model with a spinor gauge invariance at $m^2=-M^2$ has $\nu=3/2$,
just as for linearized cosmological supergravity.
\end{enumerate}
All the above fermionic partially massless theories propagate on
the null cone\footnote{The argument of~\cite{Deser:1983tm} shows
that spin~3/2 cannot take the conformal value $\nu=1/2$. This does
{\it not} imply that strictly massless spin~3/2 propagates
off-cone, since null propagation is achieved there by $\nu=3/2$.
This null propagation was already proven in the second paper of
~\cite{Townsend:1977qa}.}. It is then reasonable here also to
assume that all partially massless fermions propagate on the null
cone. The spin~$s$ theory with helicities
$(\pm(n+1/2),\ldots,\pm1/2)$ removed appears at \be
 m^2=-M^2\,(n+1)^2\,
 \ee
 with Bessel index $\nu=3/2+n$. This has also been established in
\cite{russian}. The Hilbert space of these theories is unitary in
dS, where their actions are not hermitean. In particular, the
strictly massless theory with $n=s-3/2$ occurs at
$m^2=-M^2\,(s-1/2)^2$ for $s>1/2$: Its Hilbert space is unitary
for any $\L$, but hermiticity of its action is lost in dS. The
conformal value $\nu=1/2$ is only attained by the $m=0$ spin~1/2
theory.

\vspace{.4cm}

\renewcommand{\theequation}{7.\arabic{equation}}
\setcounter{equation}{0}

\section{Massless (Dis)continuities in the Newtonian Limit}

\vspace{.4cm}

As promised in the Introduction, we outline (following [5]) the
effect of the vDVZ discontinuity in the Newtonian limit in the
source-source interactions induced by massive (as $m\rightarrow
0$) {\it vs.} strictly massless tensor fields.

%following from Newtonian.tex

The presence of a second dimensional constant provides alternative
paths, and outcomes, for the limit. In particular, the spin 2 case
with, say, two (background covariantly conserved ) sources $ (
T_{\mu\nu}, t^{\mu \nu}) $ leads to the Born exchange interaction,
\be
 I = G_{\Lambda,m} \int d^4 x \left \{  T_{\mu \nu}\,
{\cal{D}}\, t^{\mu\nu} - { m^2 - \Lambda \over 3m^2 - 2 \Lambda}
T_{\mu}\,^\mu \, {\cal{D}}\, t_{\nu}\,^\nu \right \},
\label{interaction} \ee
 where $ {\cal{D}}$ is the usual massive
$(A)dS$ scalar propagator whose $m=0$ and $\Lambda =0$ limits are
smooth and $G_{\Lambda,m}$ is the gravitational constant for the
particular $(\Lambda,m)$ model. The old
discontinuity~\footnote{The effect of $1/3$ versus $1/2$ was a
finite discrepancy between predictions for experiments involving
only slow ($t_{\mu\nu} \rightarrow t_{00}$ only) and those
involving light-like ({\it{e.g.}} $t_\mu\,^\mu = 0 $) sources.
For, and  only for, the value $1/2$ could both light bending and
Newtonian gravity agree with observation since the coupling
constant $G_{\Lambda, m}$ is used up to fix the latter's
strength.} at $\Lambda =0$ led to a relative coefficient $1/3$ in
the second term versus $1/2$ if $m^2$ is identically zero. When
$\Lambda \ne 0$, there is an infinite number of limits available;
in particular $m^2 \rightarrow 0$ followed by  $\Lambda
\rightarrow 0$ reproduces the 1/2 factor.

Before considering the details, we argue physically that the
Newtonian limit of (\ref{interaction}) must be immune to
discontinuities because, by its very definition, it is only valid
for $c \rightarrow \infty$. Thus {\it{only}} ( $ T_0\,^0 = \rho,
t_0\,^0 = \sigma$ ) fail to vanish: we have an effective scalar
theory with only slow sources and one ``experiment'' to fit with
one coupling constant. There is no ``light-bending'' to fit, as
there is no light.

If $\Lambda = 0$, the interaction is
 \be I_{0,m} \sim {2\over
3}G_{0,m}\, \int d^3 x \,\rho\, Y\, \sigma \, , \label{yukawa}
 \ee
where $Y$ is the Yukawa potential and $2 G_{0,m}/3$ is tuned to
the observed Newtonian constant. Since the Yukawa potential
reduces continuously to  $1/r$, the $m \rightarrow 0$ process is
perfectly smooth.

If, on the other hand, $\Lambda \ne 0$, the effective interaction
becomes
 \be I_{\Lambda,m} \sim G_{\Lambda,m}\, ( 1- {m^2 -\Lambda
\over 3 m^2 - 2\Lambda} )\, \int d^3 x \,\rho\, Y_\Lambda\,
\sigma\, ,
 \ee
where $Y_\Lambda$ is the generalized static Yukawa potential when
$\Lambda \ne 0$. Thus ``Newton's constant'' is
 \be G_{N}
=G_{\Lambda,m}\, {2 m^2- \Lambda \over 3 m^2 - 2\Lambda }.
\label{fraction} \ee

This $(m^2,\Lambda)$ dependence of $G_{N}$ would seem to involve
some dangerous ranges and points. However, in the original Lorentz
invariant domain whose limit this is, all models with $ 0 < 3 m^2
< 2 \Lambda $ are, as we saw earlier, non-unitary and so
unphysical. This excludes the region where the fraction in
(\ref{fraction}) would turn negative, as well as the point $2m^2 =
\Lambda$ where the numerator vanishes~\footnote{We emphasize that,
at the Newtonian level, this vanishing is a simple case of
cancellation between Newtonian attraction and the non-unitary
helicity zero ghost's repulsive contribution. It corresponds to
the covariant interaction $ (T_{\mu\nu}t^{\mu\nu}-
T_{\mu}^{\mu}t_{\nu}^{\nu}) $ in which there is manifestly no
$T_{00}t^{00}$ term.}. The $3m^2 = 2\Lambda$ model
\cite{Deser:1983tm} is unitary but has a gauge invariance that
requires its conserved sources to be traceless as well, so it has
no Newtonian limit at all. The physical region relevant to
(\ref{fraction}) thus consists of the usual gauge point $m^2 = 0$,
together with that part of the ($m^2, \Lambda$) plane for which
$m^2 > 2\Lambda/3$, including of course $AdS$ space where $\Lambda
< 0$. Any limit of $(m^2, \Lambda )\rightarrow 0$ in this region
is perfectly smooth, with a well-defined positive $G_{N}$.

\renewcommand{\theequation}{8.\arabic{equation}}
\setcounter{equation}{0} \vspace{.4cm}

\section{A Cosmological Speculation}

\vspace{.4cm}

We conclude with a cosmological consequence of having infinite
towers of higher spin systems. Previous attempts to render $\L$
small or vanishing have been of two types: (i)~Those based on
(unbroken) symmetries, such as the cancellation of supersymmetric
zero point energies~\cite{Zumino:1975bg} or the necessary absence
of a cosmological terms in $d=11$
supergravity~\cite{Bautier:1997yp}. (ii) A dynamical solution
based on quantum gravity loop corrections driving $\L$ to
zero~\cite{Tsamis:1993sx}. Our idea is rather different: it
depends only on the kinematics of a tower of free fields. Such a
leap of faith should not be foreign to string theorists, since
massive string states couple an infinite tower of higher spins.

The argument is that the region of the $(m^2,\L)$ plane where the
entire tower has unitary content only is squeezed onto the $\L=0$
axis: The unitary region for massive higher spins in the
$(m^2,\L)$ plane is bounded below in AdS by the (highest) strictly
massless fermionic line of (6.19)
 \be m^2=-\frac{\L\,(s-1/2)^2}{3}\, ,
  \ee{
 and above
by the (lowest) partially massless bosonic gauge line (the one
excluding helicity 0), of (6.11) \be m^2=\frac{\L\,s(s-1)}{3}\, .
\ee
 Therefore the unitarily allowed region is pinched around
$\L=0$ for large values of $s$ (as is manifest in our Figure). Of
course the robustness of this mechanism will be challenged by the
usual array of difficulties introduced by interactions.

\vspace{.4cm}

\section{Conclusions}

\vspace{.4cm}

I have discussed in some detail the kinematical effects of the
simplest nontrivial -- constant curvature -- gravitational
backgrounds on matter.  The initial surprise that having $\L\neq
0$ can so greatly alter the most elementary and fundamental
properties of free fields, such as gauge invariance, masslessness,
and helicity content.  Upon further reflection, one might rather
wonder that any close correspondence between the two worlds of
flat and constant curvatures should remain at all, given for
example the transmutation of Poincar\'{e} to homogeneous (A)dS
algebras. Indeed, there are bound to be more such surprises, and
of course global problems are guaranteed to exist!  Still, this
complex of ideas will certainly be of relevance to such realms as
string theory and its slope expansion, and perhaps even to the
cosmological problem's resolution.

\vspace*{.2in}

I thank A.\ Waldron and B.\ Tekin for stimulating collaborations.
This work was supported by NSF Grant PHY99-73935.

\end{document}